\documentclass[sigconf, authorversion, nonacm]{acmart}

\newcommand{\SysTitle}{RepoDoc\xspace}

\newcommand{\Sys}{RepoDoc\xspace}
\newcommand{\Bench}{RepoDocBench\xspace}
\newcommand{\RA}{RepoAgent\xspace}
\newcommand{\CW}{CodeWiki\xspace}

\newcommand{\MetCov}{Coverage\xspace}
\newcommand{\MetComplK}{Completeness@K\xspace}
\newcommand{\MetTQS}{TQS\xspace}
\newcommand{\MetUR}{Update Recall\xspace}
\newcommand{\MetDI}{Doc Information\xspace}

\newcommand{\KGATitle}{Repository Knowledge Graph\xspace}
\newcommand{\KGA}{repository knowledge graph\xspace}
\newcommand{\KG}{knowledge graph\xspace}
\newcommand{\RepoKG}{RepoKG\xspace}
\newcommand{\SIP}{Semantic Impact Propagation\xspace}
\newcommand{\SAgentTitle}{Skill-based agent\xspace}
\newcommand{\SAgent}{skill-based agent\xspace}

\newcounter{finding}
\newcommand{\finding}[2]{%
  \refstepcounter{finding}%
  \begin{tcolorbox}[
    enhanced,              
    frame hidden,          
    boxrule=0pt,           
    borderline west={3pt}{0pt}{#1}, 
    colback=#1!15,         
    left=2pt,              
    right=0pt,             
    top=2pt,               
    bottom=2pt,            
    before skip=1mm,       
    after skip=1mm         
  ]
    \textbf{Finding \arabic{finding}:} #2
  \end{tcolorbox}%
}

\AtBeginDocument{%
  }

\usepackage{tikz}
\usetikzlibrary{shapes.geometric, arrows, positioning, fit, calc}
\usepackage{minted}
\usepackage{pgfplots}
\pgfplotsset{compat=1.17}
\usepgfplotslibrary{groupplots}
\usepackage{svg}
\usepackage{multirow}
\usepackage{diagbox}
\usepackage{algorithm}
\usepackage{algorithmicx}
\usepackage{algpseudocode}
\usepackage{graphicx}
\usepackage{subcaption}
\usepackage{url}
\usepackage{xspace}
\usepackage[most]{tcolorbox}

\renewcommand\footnotetextcopyrightpermission[1]{}
\begin{document}

\title[\SysTitle: Documentation via \KGATitle]{RepoDoc: A Knowledge Graph-Based Framework to Automatic Documentation Generation and Incremental Updates}

\author{Dong Xu}
\affiliation{%
  \institution{Sun Yat-sen University}
  \country{China}
}
\email{xudong7@mail2.sysu.edu.cn}

\author{Mingwei Liu}
\authornote{Corresponding author.}
\affiliation{%
  \institution{Sun Yat-sen University}
  \country{China}
}
\email{liumw26@mail.sysu.edu.cn}

\author{Xiwen Wang}
\affiliation{%
  \institution{Sun Yat-sen University}
  \country{China}
}
\email{wangxw86@mail2.sysu.edu.cn}

\author{Jianfeng Zhong}
\affiliation{%
  \institution{Sun Yat-sen University}
  \country{China}
}
\email{zhongjf25@mail2.sysu.edu.cn}

\author{Zibin Zheng}
\affiliation{%
  \institution{Sun Yat-sen University}
  \country{China}
}
\email{zhzibin@mail.sysu.edu.cn}

\renewcommand{\shortauthors}{Xu et al.}

\begin{abstract}
Maintaining up-to-date, comprehensive documentation for large codebases is a persistent challenge. Recent progress in automated documentation has moved from template-based rules to large language models (LLMs), yet existing tools still process source code as flat fragments, producing isolated documents that lack semantic structure. This design also leads to excessive token consumption and slow generation, while failing to capture how code changes propagate across dependencies.

We propose RepoDoc, a system that uses a repository knowledge graph (RepoKG) as the semantic foundation for the entire documentation lifecycle. Our framework consists of three stages: (1) RepoKG construction, which extracts code entities and their relationships; (2) module clustering, which groups code into functionally cohesive, hierarchical units; and (3) skillful agent-based generation, which queries the graph to create modular, cross-referenced documentation with auto-generated Mermaid diagrams. For incremental maintenance, a semantic impact propagation mechanism navigates the RepoKG bidirectionally to pinpoint all affected parts, allowing selective, targeted regeneration.

Evaluated on 24 repositories across 8 programming languages, RepoDoc substantially outperforms state-of-the-art alternatives. It improves API coverage by 32.5\% and completeness by 10.4\%, while generating documentation 3× faster with 85\% fewer tokens. For incremental updates, it cuts update time by 73\% and token usage by 77\%, and achieves 10.2\% higher update recall, more accurately reflecting code changes in the regenerated documentation.

The source code and experimental artifacts are available at \url{https://github.com/SYSUSELab/RepoDoc}.
\end{abstract}





\maketitle

\section{Introduction}
High-quality documentation is essential for software maintenance and comprehension in long-evolving projects \cite{allamanis2018surveymachinelearningbig,sym14030471}. Without adequate documentation, developers face significant challenges in understanding and navigating codebases \cite{ahmad2020transformerbasedapproachsourcecode,10.1145/3180155.3180167}. Studies show that developers spend up to 58\% of their time understanding code rather than writing it \cite{xia2017measuring}, and inadequate documentation further increases this cognitive burden \cite{ahmad2020transformerbasedapproachsourcecode,10.1145/3180155.3180167}. However, manually writing and maintaining comprehensive documentation is both time-consuming and error-prone \cite{allamanis2016convolutionalattentionnetworkextreme,allamanis2018surveymachinelearningbig,wang2021codet5identifierawareunifiedpretrained}. As codebases evolve, documentation frequently becomes stale, requiring substantial effort to keep it synchronized with code changes \cite{hou2024largelanguagemodelssoftware,sym14030471}. These pervasive challenges have motivated extensive research into automated documentation generation \cite{luo2024repoagent,makharev2025codewiki}, yet current tools still struggle to produce accurate, modular, and well-structured documentation at scale \cite{gao2023rag}.

The landscape of automated documentation has evolved significantly \cite{zhang2024unifyingperspectivesnlpsoftware,10.1145/3747588,inproceedings222}. Early approaches relied on rule-based templates and statistical models for generating individual function comments \cite{10.1145/3747588,allamanis2016convolutionalattentionnetworkextreme}. The emergence of deep learning, particularly transformer-based pre-trained models such as CodeBERT \cite{codebert2020} and GraphCodeBERT \cite{guo2021graphcodebertpretrainingcoderepresentations}, advanced code understanding through semantic representations. Large Language Models (LLMs) further expanded these capabilities with enhanced code understanding, long-range context modeling, and coherent long-form generation, enabling repository-level documentation generation that was previously infeasible \cite{guo2024deepseekcoderlargelanguagemodel,li2023starcodersourceyou,gao2023rag}. Among contemporary LLM-powered tools, \RA \cite{luo2024repoagent} constructs a bidirectional reference graph for documentation ordering, while \CW \cite{makharev2025codewiki} proposes hierarchical decomposition for large-scale codebases. Despite these advances, existing approaches still treat code as flat text chunks or physical file structures, resulting in isolated documentation without semantic modularity. \textbf{Furthermore, these approaches often rely on extensive code retrieval to assemble context, which floods the LLM context with excessive code, leading to severe token inflation (e.g., 5.3K tokens on average for \CW) and prohibitively slow generation (e.g., 3,673 seconds per repository), making practical deployment challenging.} Moreover, they lack the semantic understanding required to track how code changes propagate through dependencies, rendering incremental updates inaccurate or entirely unsupported \cite{luo2024repoagent,makharev2025codewiki}.

\begin{figure*}[t]
\centering
\includegraphics[width=1\linewidth]{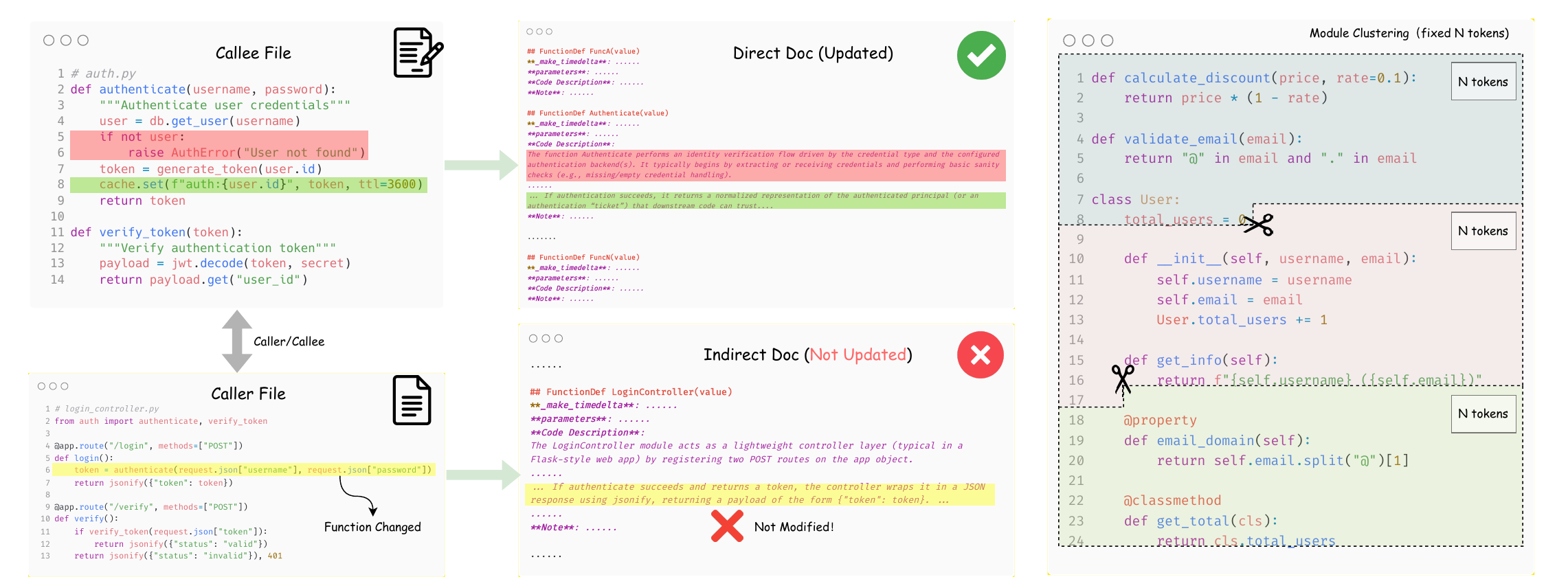}
\caption{Baseline problems: \textbf{Left:} \RA uses physical file structure with template-based documentation; \textbf{Right:} \CW decomposes code by physical size}
\label{fig:baseline-problems}
\end{figure*}

To address these limitations, we present \textbf{\Sys}, an automated documentation generation system that leverages \KGA(\RepoKG) as the semantic backbone for the entire documentation lifecycle. Unlike existing methods that rely on physical file structures \cite{luo2024repoagent,makharev2025codewiki}, our core insight is to \textit{construct a \RepoKG once, and leverage it as a semantic backbone to orchestrate both context-aware document generation and precise incremental updates}. \Sys employs a three-phase pipeline. First, \textit{\KGA construction} extracts code entities and their semantic relationships (calls, implements, imports, etc.) to construct a \RepoKG, overcoming the limitations of flat file structures by enabling semantic-aware organization. Second, \textit{module clustering} leverages the \RepoKG to group code into hierarchical structures based on functional cohesion rather than physical layout, yielding well-structured documentation with clear architectural overviews. Third, \textit{documentation generation} employs a \SAgent architecture. By utilizing precise graph queries, the agents retrieve only relevant code snippets rather than entire files, thereby avoiding context flooding and accelerating generation. For incremental updates, \Sys introduces a \SIP mechanism that traverses the \RepoKG bidirectionally to identify all affected components, enabling the selective regeneration of only the changed documentation. \textbf{Consequently, \Sys drastically reduces token consumption and generation time while ensuring highly accurate documentation maintenance during code evolution.}

To enable fair and comprehensive evaluation, we develop \textbf{\Bench}, a benchmark framework featuring five evaluation dimensions: \MetCov, \MetDI, \MetComplK, \MetTQS, and \MetUR. \Bench encompasses 24 repositories across 8 programming languages, covering diverse scales from small libraries (<10K LOC) to large enterprise codebases (>100K LOC). It also introduces a novel three-document comparison mechanism to rigorously assess the quality and accuracy of incrementally updated content.

Experiments on \Bench demonstrate that \Sys significantly outperforms state-of-the-art baselines. Specifically, \Sys achieves 32.5\% higher API coverage, meaning more public APIs are documented. It attains 10.4\% higher completeness, meaning the documentation provides more comprehensive descriptions of code functionality. Generation is 3$\times$ faster with 85\% token reduction, yielding substantial savings in both time and cost. For incremental updates, \Sys reduces update time by 73\% and token usage by 77\%. Moreover, it achieves 10.2\% higher update recall, correctly capturing more code changes in the updated documentation.

The key contributions of this paper are:
\begin{enumerate}
    \item \textbf{\Sys System}: A novel documentation generation framework that employs \RepoKG to capture code dependencies. Coupled with a \SAgent pipeline, \Sys produces modular, cross-referenced documentation with auto-generated architecture diagrams, and features \SIP for precise incremental updates.
    \item \textbf{\Bench}: A comprehensive benchmark framework covering diverse languages and scales, equipped with four multi-dimensional metrics and a rigorous three-document comparison mechanism for fair incremental update assessment.
    \item \textbf{Extensive Evaluation}: Comprehensive experiments on 24 real-world repositories demonstrating that \Sys significantly outperforms state-of-the-art baselines in both generation quality, computational efficiency, and incremental update accuracy.
\end{enumerate}

\section{Motivation}
\label{motivation}

\begin{figure*}[t]
\centering
\includegraphics[width=1\linewidth]{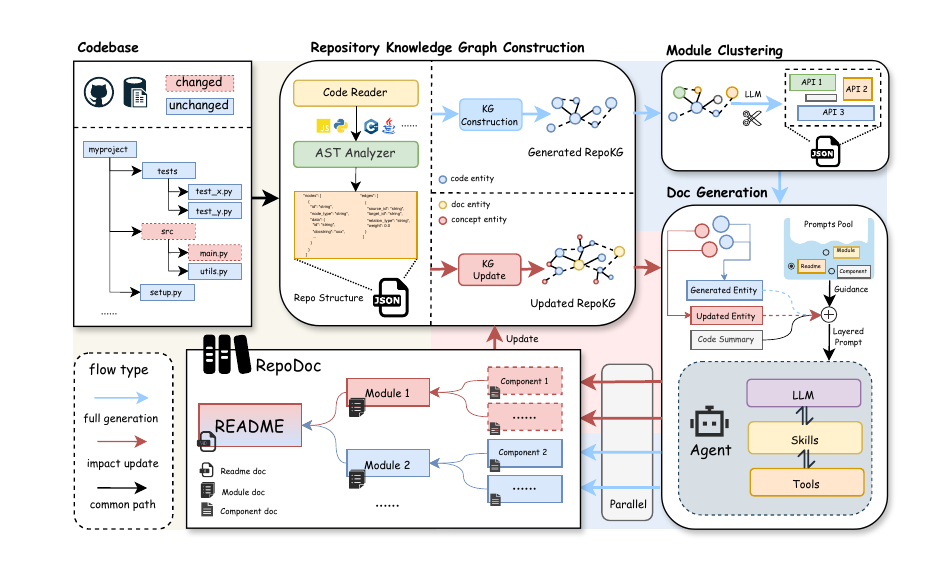} 
\caption{\Sys architecture. Blue arrows: full generation workflow; Red arrows: incremental update workflow; Black arrows: common paths.}
\label{fig:architecture}
\end{figure*}

Despite the promising results of recent LLM-based approaches, critical limitations remain in practical deployment. To understand these limitations, we analyze two representative systems: \RA \cite{luo2024repoagent} and \CW \cite{makharev2025codewiki}. As the most recent state-of-the-art frameworks, they perfectly capture the current landscape of LLM-powered repository-level documentation—one relying on AST-driven physical file orchestration (\RA) and the other on token-limited hierarchical decomposition (\CW). Our analysis reveals three fundamental weaknesses that limit their effectiveness in real-world documentation scenarios.

\noindent\textbf{Flat and Isolated Documentation}. \RA organizes documentation based on the physical file structure of the original codebase, which fails to capture semantic relationships across modules. The generated documentation appears as a collection of isolated pages with rigid, template-based structure (e.g., FunctionDef, function name, parameters, code description, note), lacking meaningful cross-references or functional reorganization. \CW, while employing hierarchical decomposition, relies on physical code size rather than semantic coherence—the recursive module splitting is determined by token counts and character limits rather than actual functional cohesion. This results in fragmented documentation chunks that lack structural integration and fail to provide a coherent view of the codebase architecture.

\noindent\textbf{Excessive Token and Time Costs}. Both tools adopt retrieval strategies that flood the LLM context with excessive code. \RA injects all caller-callee relationships into the prompt, which can easily reach tens of thousands of tokens for frequently-used components in large repositories. \CW's bottom-up synthesis mechanism requires parent nodes to absorb complete source code, all sub-module documentation, and structural/dependency information—a form of context stacking that leads to exponential token inflation. This results in prohibitively long generation times and high API costs.

\noindent\textbf{Limited Incremental Update Capabilities}. For code evolution scenarios, \RA only detects file-level changes without semantic understanding—it cannot trace how an interface modification affects implementing classes or calling functions through indirect relationships. \CW lacks any incremental support entirely, requiring full regeneration for every code change. This makes documentation maintenance computationally prohibitive for large repositories and risks losing previously generated context. Figure \ref{fig:baseline-problems} illustrates these baseline problems in detail.

These observations motivate our insight: leveraging \RepoKG as the semantic backbone for the documentation lifecycle. Unlike the flat or physically-decomposed structures used by existing approaches, \RepoKG naturally provides three key capabilities. First, it enables semantic-aware organization that transcends physical file boundaries. Second, it allows precise context retrieval that avoids flooding the LLM with excessive code. Third, it supports comprehensive impact analysis that can trace how code changes propagate through the system. This insight draws from the successful application of knowledge graphs in software engineering tasks such as code search, bug detection~\cite{zhou2019devigneffectivevulnerabilityidentification}, and program comprehension \cite{wang2020survey}, as well as theoretical foundations in change impact analysis~\cite{ryder2001change}, refactoring detection~\cite{10.1145/3180155.3180206} and fine-grained source code differencing \cite{falleri2014fine}.

\section{Approach}
\label{approach}

\begin{figure}[htbp]
\centering
\includegraphics[width=1\linewidth]{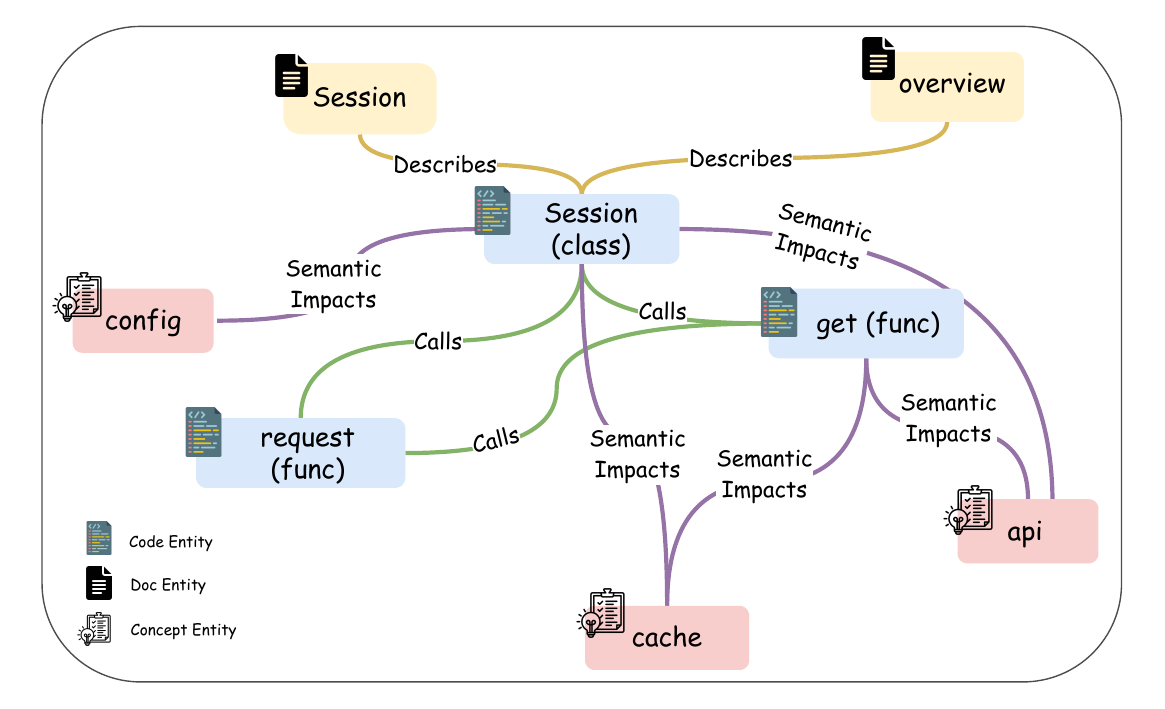}
\caption{Example of \RepoKG structure}
\label{fig:kg-structure}
\end{figure}

Motivated by the challenges and insights discussed in the previous section, we present \textbf{\Sys}, a \KG-based documentation generation system that takes a source code repository as input and produces comprehensive multi-level documentation including README, module documentation, and API reference documentation. As code evolves, \Sys also supports incremental updates that selectively regenerate only affected documentation components rather than regenerating everything from scratch. Unlike baseline approaches that rely on flat text chunks or physical file structures, \Sys builds a \RepoKG as the semantic backbone, which serves as the central data structure throughout the entire documentation lifecycle. Figure \ref{fig:architecture} illustrates the overall architecture of our system.

As illustrated in Figure \ref{fig:architecture}, \Sys employs a three-phase pipeline. Our approach centers on building a \RepoKG as the semantic backbone, which enables three key capabilities. First, it organizes code into hierarchical module structures using \RepoKG analysis and LLM-powered clustering for well-structured documentation. Second, it generates comprehensive documentation using a \SAgent architecture with precise graph queries to control token consumption. Third, it uses \RepoKG traversal to trace \SIP for efficient incremental updates.

\subsection{\KGATitle Construction}
\label{dependency}

The first phase constructs a \KGA as the backbone for documentation generation. Given a source code repository as input, the phase outputs a \KGA capturing code entities and their relationships.

\subsubsection{\RepoKG Schema}
We define the \RepoKG schema with three entity types and seven relationship types. Code Entity represents code constructs (functions, classes, interfaces, modules) along with source code, signatures, metadata, and their dependencies to other code entities. Concept Entity represents business concepts extracted via LLM analysis, capturing domain knowledge not explicit in code structure. Doc Entity stores generated documentation content in markdown format, enabling selective updates and cross-reference navigation between documents. The seven relationship types include calls (function invocations), implements (class-interface), extends (inheritance), imports (module dependencies), contains (module containment), semantic-impact (code-to-concept), and describes (documentation-to-code).

\subsubsection{Construction Process}
The construction process consists of three sequential steps.

\textbf{Code Entity Extraction.} The source code is parsed using appropriate tools (native AST for Python, tree-sitter for other languages) to extract functions, classes, interfaces, and modules as Code Entity instances. Each entity captures the source code, signature, and metadata such as file path and visibility modifiers.

\textbf{Structural Relationship Extraction.} Four types of structural relationships are derived directly from code structure without semantic analysis. Imports relationships are extracted from import statements. Contains relationships are derived from file system hierarchy and namespace analysis. Extends and implements relationships are parsed from class inheritance declarations. Calls relationships are obtained via intra-procedural control flow analysis. Together, these form the structural skeleton of the \RepoKG.

\textbf{Semantic Relationship Enrichment.} The structural skeleton is augmented with semantic relationships using LLM-based analysis. For each Code Entity, the LLM identifies relevant business concepts, creating Concept Entity instances linked via semantic-impact relationships. During documentation generation, describes relationships are established between generated documentation and the Code Entity it documents.

\begin{figure}[t]
\centering
\includegraphics[width=1\linewidth]{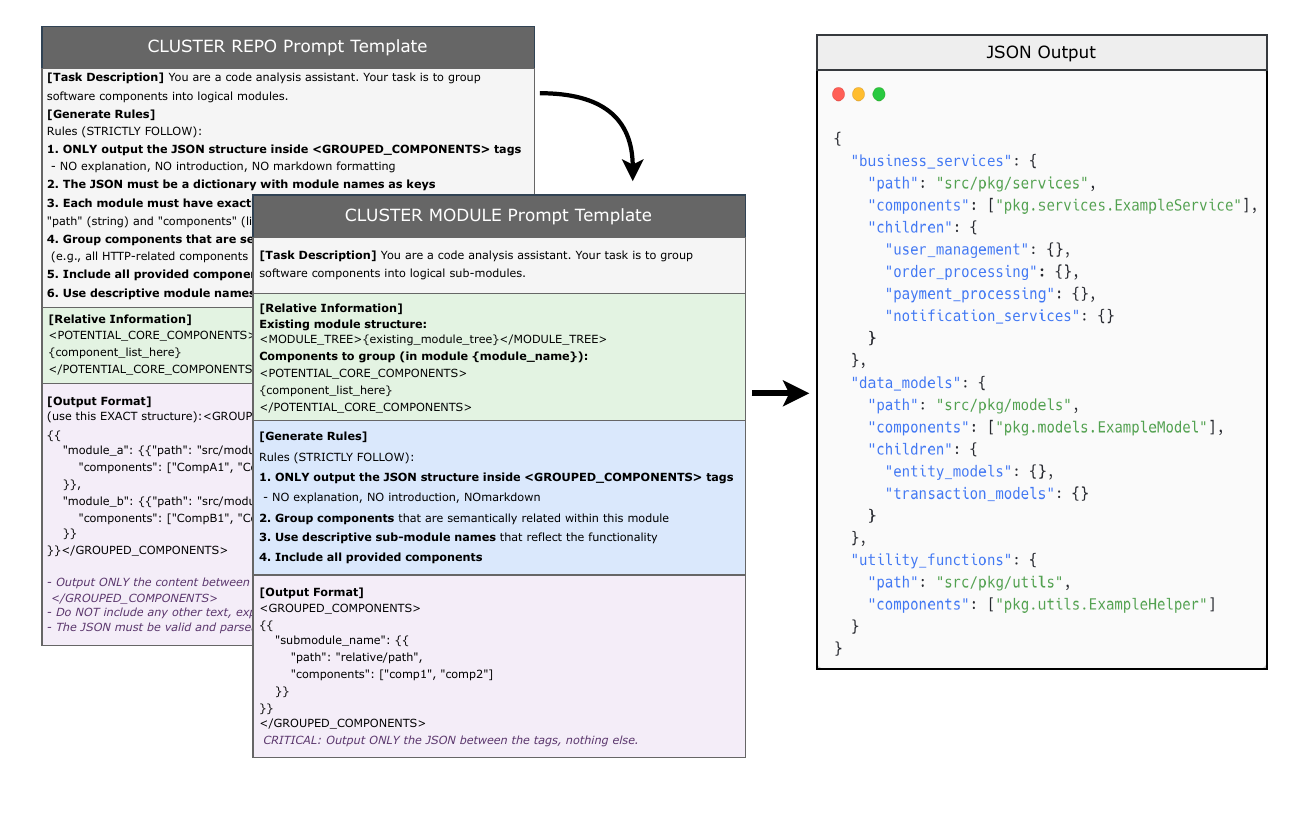}
\caption{Clustering prompt and output example}
\label{fig:clustering-prompt}
\end{figure}

\subsection{Module Clustering}
\label{clustering}

The second phase partitions code entities into a hierarchical module structure. This hierarchy is stored persistently in the \RepoKG as Module Entity instances, enabling efficient queries during documentation generation and incremental updates.

\textbf{Input and Output.} Given the \RepoKG from Phase 1, the phase outputs a tree-structured hierarchy of Module Entity instances, where each Module Entity contains references to its member Code Entity instances via contains relationships.

\textbf{Recursive top-down clustering.} The algorithm employs a recursive top-down strategy. For a given set of code entities, the LLM partitions them into $K$ coherent modules based on functional cohesion and semantic relatedness, leveraging the structural relationships (calls and imports) from the \RepoKG. If a resulting module exceeds a token threshold, its sub-components are recursively clustered until all modules satisfy the token limit.

In our implementation, top-level clustering uses $K=5$, and recursive sub-clustering uses $K=3$ with a threshold $T$ of 4,096 tokens. Clustering leverages call and import relationships to inform grouping decisions---entities with dense inter-connections are preferentially grouped.

Figure \ref{fig:clustering-prompt} illustrates the clustering prompt and output example.

\subsection{Documentation Generation}
\label{generation}

The final phase generates comprehensive documentation using AI agents. Given the \RepoKG and module structure as input, the phase outputs multi-level documentation including README, module docs, and component API docs.

\begin{figure}[htbp]
\centering
\includegraphics[width=1\linewidth]{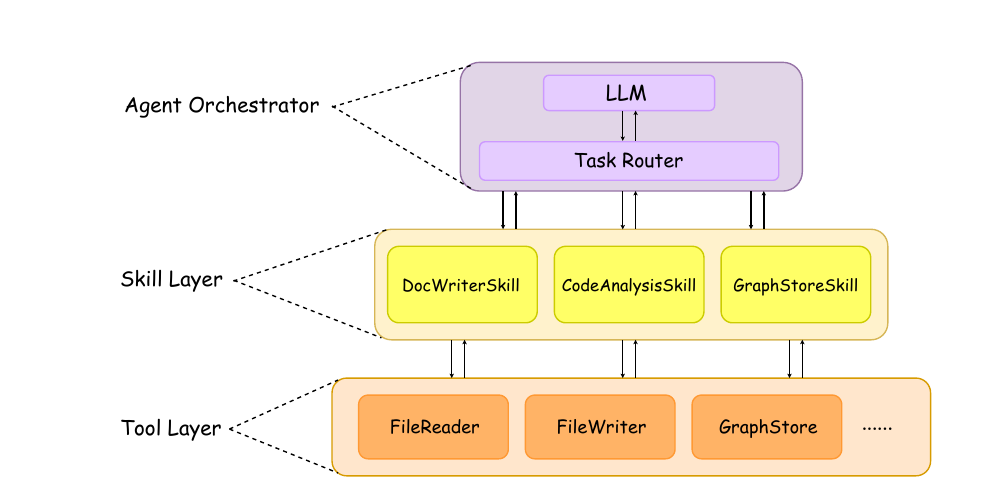}
\caption{\SAgentTitle architecture}
\label{fig:skill-architecture}
\end{figure}

As illustrated in Figure \ref{fig:skill-architecture}, the implementation employs a layered agent architecture consisting of an orchestrator, a skill layer, and a tool layer.

\textbf{Agent Orchestrator.} The orchestrator comprises an LLM and a task router. The LLM serves as the brain, responsible for understanding instructions, reasoning, and planning~\cite{hong2024metagptmetaprogrammingmultiagent,yao2023reactsynergizingreasoningacting}. The task router decomposes high-level plans from the LLM and dispatches tasks to appropriate skills. This design addresses the limitations of traditional single-agent approaches that rely on monolithic prompting, which floods the LLM with excessive context.

\textbf{Skill Layer.} The skill layer contains three specialized skills. The documentation writer generates API references with parameter tables, method descriptions, usage examples, and Mermaid diagrams for module structure. The code analyzer extracts business concepts and analyzes dependencies during generation to enrich the \RepoKG. The graph query skill retrieves relevant context about related components and dependencies during documentation generation.

\textbf{Tool Layer.} The tool layer provides atomic operations for the skills. FileReader accesses source code. FileWriter saves generated documentation. GraphStore manages \RepoKG operations. LLMCaller invokes the language model for content generation.

\textbf{Generation Strategy.} The generation process follows a bottom-up approach: component-level documentation is generated first, then module-level documentation that references component docs, and finally the repository overview that synthesizes the entire documentation structure. Components are processed in parallel using thread pools, respecting dependency relationships. Module documentation generation respects the hierarchy by processing leaf modules first and progressively building toward top-level summaries.

\subsection{Incremental Update}
\label{incremental}

When code evolves, documentation must be updated accordingly to remain accurate. Existing documentation generation tools typically require full regeneration when code changes, which is computationally expensive and may lose previously generated context. The key challenge lies in determining the precise scope of impact: a single line change in a widely-used API may cascade through thousands of dependent components, while localized changes to internal implementation details may have minimal reach. Simply regenerating all documentation wastes computational resources, yet missing affected components produces inconsistent or incorrect documentation.

Our incremental update mechanism takes the existing \RepoKG, documentation, and a commit diff (representing code changes between two commits) as input, and produces updated documentation with minimal regeneration. It consists of four sequential stages: change detection identifies modified code entities via AST analysis and classifies change types; \SIP determines affected components through \RepoKG traversal; selective regeneration updates only affected documentation while preserving unchanged content; and validation ensures cross-reference consistency across the documentation hierarchy.

\subsubsection{Change Detection}
The first stage identifies what has changed in the codebase. Git diff analysis detects modified, added, and deleted files between commits. For each changed file, the AST is parsed to determine which code entities have been affected, and the change type is classified to determine the appropriate update strategy. Changes are categorized based on their impact scope: additions and signature modifications require new documentation generation, removals trigger documentation cleanup, while body changes may propagate to dependent components based on visibility levels.

The output of this stage is a list of changed files along with the affected code entities and their change classifications.

\subsubsection{\SIP}
The second stage determines which components are affected by the detected changes. Code changes in one component often affect dependent components through semantic relationships. To determine the propagation scope, the \RepoKG is traversed in both directions: downstream (components that depend on the changed code) and upstream (components that the changed code depends on). Starting from the changed components, both their dependents and their dependencies are iteratively explored.

\textbf{Different relationship types inform propagation decisions with varying implications.} Our \RepoKG captures multiple relationship types: direct call relationships propagate changes broadly through the call chain; interface implementation relationships ensure consistency across polymorphic dependencies; semantic impact relationships track non-obvious couplings such as shared data structures or contract violations that may require documentation updates. Based on the change type detected in the previous stage, we determine the appropriate propagation scope.

\textbf{Bidirectional traversal discovers affected components efficiently.} As illustrated in Algorithm \ref{algo:impact-propagation}, starting from the initially changed components $C$, the algorithm iteratively discovers affected components by traversing both downstream edges (successors) and upstream edges (predecessors). For each visited component, we examine its relationships—if the relationship type warrants propagation, the connected component is added to both the affected set $A$ and the processing queue. This continues until all reachable components have been visited, naturally discovering components in order of their distance from the original change.

\begin{algorithm}[h]
\caption{\SIP}
\label{algo:impact-propagation}
\begin{algorithmic}
\State \textbf{Input:} Changed components $C$, \RepoKG $G$
\State \textbf{Output:} Affected component set $A$
\Function{SemanticImpactPropagation}{$C$, $G$}
    \State $A \gets C$
    \State $Queue \gets C$
    \While{$Queue \neq \emptyset$}
        \State $v \gets \text{dequeue}(Queue)$
        \ForAll{$s \in \text{connected}(v, G)$}
            \If{$\text{relationship}(v, s)$ \text{is propagation-worthy}}
                \State $A \gets A \cup \{s\}$
                \State $\text{enqueue}(Queue, s)$
            \EndIf
        \EndFor
    \EndWhile
    \State \Return $A$
\EndFunction
\end{algorithmic}
\end{algorithm}

\textbf{Topological sorting ensures consistent documentation updates.} To ensure documentation consistency, we use Kahn's algorithm to compute a topological ordering of affected components $A$ based on their dependency relationships. This algorithm first identifies root nodes (components with no incoming edges) and progressively processes them, adding their dependents and maintaining the order. The result is a processing sequence where each component appears before all components that depend on it, ensuring that lower-level components are updated before higher-level ones that reference them. The propagation stage typically produces a set of affected components that represents a small fraction of the entire codebase for localized changes.

\subsubsection{Selective Regeneration}
The third stage updates only the documentation of affected components. Leveraging the change classifications from the first stage and the affected component set from the second stage, the system determines precisely what needs to be regenerated. Instead of regenerating the entire documentation, this stage selectively updates only the affected components. For each affected component, the previously generated documentation is preserved and only what is necessary is regenerated. Additionally, module-level overviews are updated and diagrams that reference affected components are rebuilt.

This strategy significantly reduces computational cost and token usage compared to full regeneration, while maintaining documentation quality and consistency.

\subsubsection{Validation}
The final stage verifies that cross-references remain consistent across the documentation hierarchy. Both rule-based validation and LLM-based verification are employed to ensure: (1) internal links point to valid sections, (2) diagram references are correct and properly formatted, and (3) module-level documentation accurately reflects the state of their constituent components. This validation ensures that incrementally updated documentation maintains the same quality and coherence as fully regenerated documentation.

\section{Evaluation}
\label{experiments}

To evaluate \Sys's effectiveness, we design a multi-dimensional evaluation framework and formulate four research questions (RQs) aligned with our contributions:

\begin{itemize}
    \item \textbf{RQ1}: How complete is the documentation generated by \Sys compared to baselines?
    \item \textbf{RQ2}: How readable and rich is the documentation generated by \Sys compared to baselines?
    \item \textbf{RQ3}: How efficient is \Sys in terms of time and token consumption compared to baselines?
    \item \textbf{RQ4}: How effectively does \Sys perform incremental updates compared to baselines?
\end{itemize}

\subsection{Experimental Setup}
\label{setup}

We set up experiments on \Bench using DeepSeek V3.2 as the generation backbone, comparing \Sys against two state-of-the-art baselines. Our evaluation covers documentation completeness (RQ1), readability and richness (RQ2), efficiency (RQ3), and incremental update performance (RQ4).

\begin{table}[t]
\small
\caption{\Bench Dataset Statistics}
\label{tab:dataset}
\centering
\resizebox{1\linewidth}{!}{
\begin{tabular}{@{}ccc@{}}
\toprule
\textbf{Language} & \textbf{Repositories} & \textbf{Scale} \\
\midrule
Python & requests, flask, pytest, fastapi & S/M/L \\
JS/TS & axios, express, vue, react & M/L \\
Java & mybatis, guava, spring-boot, elasticsearch & M/L \\
C\# & aspnetcore, entityframeworkcore, xunit, moq & M/L \\
C/C++ & libevent, libuv, nghttp2, protobuf & M/L \\
PHP & guzzle, monolog, laravel, symfony & M/L \\
\bottomrule
\end{tabular}
}
\end{table}

\subsubsection{\Bench}
\label{benchmark}

To enable fair and comprehensive evaluation, \Bench consists of five evaluation metrics and a repository collection.

\textbf{\MetCov (Cov.)} measures the percentage of public APIs with corresponding documentation. It extracts all APIs from the source code via AST parsing and checks whether each API is mentioned in the generated documentation, computing: $\text{coverage} = \frac{\text{covered\_apis}}{\text{total\_apis}} \times 100\%$.

\textbf{\MetDI (Doc Info.)} evaluates documentation richness across five dimensions: words, files, cross-references, code blocks, and diagrams. These metrics assess structural elements, visualizations, and content density in the generated documentation.

\begin{figure}[t]
\centering
\includegraphics[width=1\linewidth]{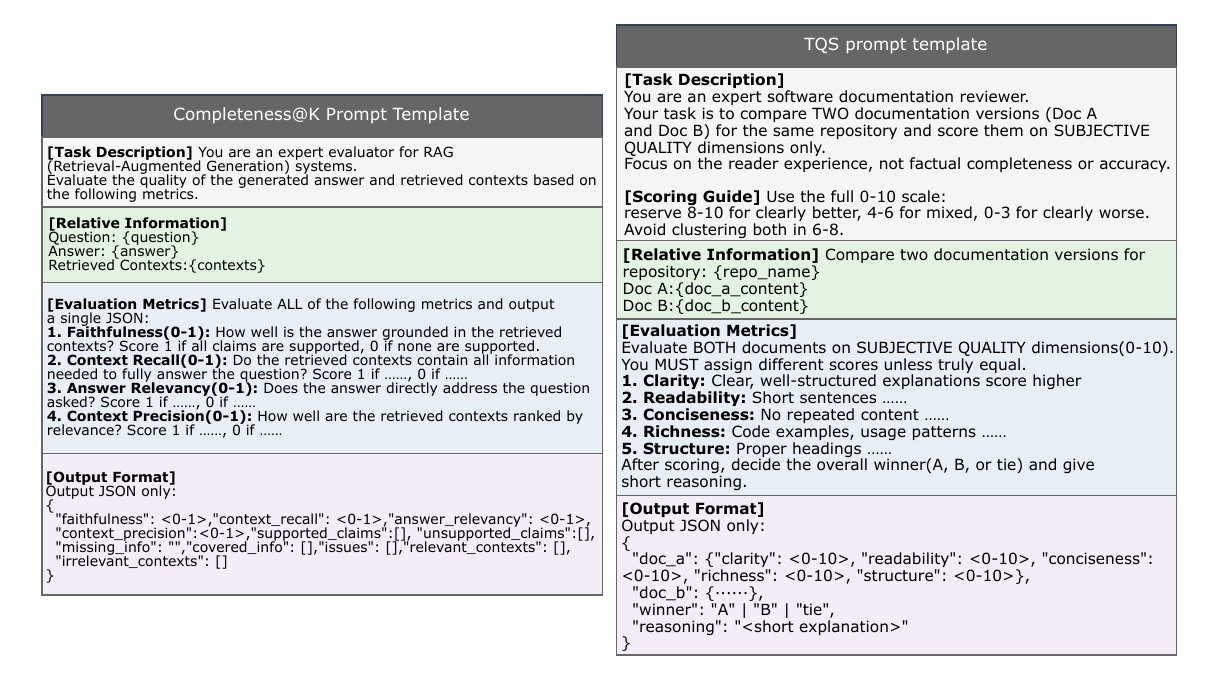}
\caption{Evaluation prompts for LLM judges: Completeness@K (left) and TQS (right)}
\label{fig:eval-prompts}
\end{figure}

\textbf{\MetComplK (Compl@K)} measures the information completeness of generated documentation. For each question, it retrieves the top-K most relevant document chunks using Sentence-BERT embeddings~\cite{reimers2019sentencebertsentenceembeddingsusing} (all-MiniLM-L6-v2), then uses an LLM judge to determine whether these chunks contain sufficient information to answer the question (see Figure \ref{fig:eval-prompts} for the evaluation prompts). The final metric is the percentage of questions that can be adequately answered: $\text{Compl@K} = \frac{\text{answerable\_questions}}{\text{total\_questions}} \times 100\%$.

\textbf{\MetTQS} uses LLM judges to compare documentation from two tools. For each comparison, the judge provides a single assessment rating all five dimensions on a 0-10 scale simultaneously: Clarity (ease of understanding), Readability (flow and coherence), Conciseness (appropriate detail without verbosity), Richness (informative content and examples), and Structure (logical organization). Position bias is mitigated by swapping document order and averaging scores.

\textbf{\MetUR (Upd Recall)} measures the proportion of code changes that are correctly reflected in the updated documentation. It answers: "Among all the components that should have been updated, how many were actually covered in the updated documentation?" Computed as: $\text{Upd Recall} = \frac{\text{correctly\_updated\_components}}{\text{total\_components\_requiring\_update}} \times 100\%$.

RQ1 uses \MetCov, \MetDI, and \MetComplK to evaluate documentation completeness. RQ2 uses \MetTQS to evaluate documentation richness. RQ3 uses Time and Token consumption to evaluate efficiency. RQ4 uses \MetUR to evaluate incremental update performance.

\textbf{Repository Collection.} We curate a diverse dataset of 24 open-source repositories across 8 programming languages (Python, JavaScript, TypeScript, Java, C\#, C, C++, PHP), classified into three scales: small ($<$10K LOC), middle (10K-100K LOC), and large ($>$100K LOC). Since \RA only supports Python, we use Python for fair three-tool comparison. RQ4 evaluates incremental updates on Python repositories only.

Table \ref{tab:dataset} summarizes the dataset. Each repository represents real-world software with varying complexity, ensuring comprehensive evaluation across different domains and scales.

\begin{table}[t]
\small
\caption{Language Support and Incremental Update Capability Comparison}
\label{tab:method-comparison}
\centering
\resizebox{1\linewidth}{!}{
\begin{tabular}{cccccccc}
\toprule
\textbf{Method} & \textbf{Python} & \textbf{JS/TS} & \textbf{Java} & \textbf{C\#} & \textbf{C/C++} & \textbf{PHP} & \textbf{Incremental} \\
\midrule
\RA & $\checkmark$ & $\times$ & $\times$ & $\times$ & $\times$ & $\times$ & $\checkmark$ \\
\CW & $\checkmark$ & $\checkmark$ & $\checkmark$ & $\checkmark$ & $\checkmark$ & $\checkmark$ & $\times$ \\
\Sys & $\checkmark$ & $\checkmark$ & $\checkmark$ & $\checkmark$ & $\checkmark$ & $\checkmark$ & $\checkmark$ \\
\bottomrule
\end{tabular}
}
\end{table}

\subsubsection{Baselines}
\label{baselines}

We compare \Sys against two state-of-the-art documentation generation tools across RQ1-RQ3. For RQ4 (incremental update evaluation), only \RA is used since \CW does not support incremental updates.

\textbf{\RA} is an LLM-powered framework for repository-level documentation with iterative refinement, supporting basic file-level incremental updates. It serves as a comparison for both initial generation and incremental update capabilities.

\textbf{\CW} uses retrieval-augmented generation for holistic documentation but does not support incremental updates, requiring full regeneration for each change. It serves as a comparison for initial generation quality only.

\subsubsection{Implementation}
\label{implementation}

We use DeepSeek V3.2 as the generation backbone for both baselines and \Sys, configured with temperature set to 0.7 and max tokens set to 4,096 for consistent generation quality. For evaluation, we use GPT-5.2 and Claude-Sonnet-4.6 as judges for assessing documentation quality. Prior work has extensively demonstrated that state-of-the-art LLMs, when acting as automated evaluators (LLM-as-a-Judge), can achieve high agreement with human judgments in both general generative tasks \cite{zheng2023judging} and specific software engineering tasks such as code summarization and documentation \cite{crupi2024effectiveness, zhao2024codejudge}. To further validate the reliability of this automated metric in our specific context, we conducted a pilot study on a subset of our dataset, where the LLM-assigned scores correlated strongly with expert human annotations (Pearson's $r > 0.85$).

\subsection{RQ1: Documentation Completeness}
\label{rq1}

This research question evaluates the documentation completeness of \Sys compared to baselines.

\subsubsection{Design}
To answer RQ1, we evaluate \Sys's documentation completeness using \MetCov, \MetDI, and \MetComplK (defined in Section \ref{benchmark}).

We run \Sys and all baselines on the 24-repository dataset using DeepSeek V3.2 as the generation backbone. Since \RA only supports Python, we evaluate all methods on Python repositories for fair comparison. \CW is evaluated on 8 languages (Python, JavaScript, TypeScript, Java, C\#, C, C++, PHP).

\subsubsection{Results}
Table \ref{tab:rq1-results} shows per-repository results across all languages.

\begin{table}
\small
\caption{Per-Repository Results across Coverage, Completeness@K, and Doc Information (DeepSeek V3.2)}
\label{tab:rq1-results}
\centering
\resizebox{1\linewidth}{!}{
\begin{tabular}{ccccccccccc}
\toprule
\multirow{2}{*}{\textbf{Language}} & \multirow{2}{*}{\textbf{Method}} & \multirow{2}{*}{\textbf{Cov.}} & \multirow{2}{*}{\textbf{Compl@10}} & \multicolumn{5}{c}{\textbf{Doc Info.}} \\
\cmidrule(lr){5-9}
    & & & & \textbf{Words} & \textbf{Files} & \textbf{Cross Refs} & \textbf{Code Blocks} & \textbf{Diagrams} \\
\midrule

\multirow{3}{*}{Python} & \RA & \textbf{91.28} & 76.53 & 935,267 & 111 & 105,625 & 1,228 & 0.0 \\
    & \CW & 49.22 & 65.14 & \textbf{34,612} & 24 & 11,255 & 4,230 & 81 \\
    & \Sys & 63.15 & \textbf{78.92} & \textbf{269,626} & \textbf{152} & \textbf{77,117} & \textbf{37,042} & \textbf{73} \\
\midrule

\multirow{2}{*}{JS/TS} & \CW & 29.05 & 48.90 & 55,692 & 37 & 18,025 & 7,724 & 124 \\
    & \Sys & \textbf{34.88} & \textbf{59.99} & \textbf{452,284} & \textbf{244} & \textbf{115,032} & \textbf{75,511} & \textbf{145} \\
\midrule

\multirow{2}{*}{Java} & \CW & 70.65 & 61.69 & 118,785 & 79 & 41,793 & 14,767 & \textbf{254}\\
    & \Sys & \textbf{87.53} & \textbf{72.05} & \textbf{394,847} & \textbf{202} & \textbf{124,583} & \textbf{54,354} & 168 \\
\midrule

\multirow{2}{*}{C\#} & \CW & 22.37 & \textbf{84.56} & 119,624 & 89 & 47,252 & 14,537 & \textbf{258}\\
    & \Sys & \textbf{37.96} & 83.35 & \textbf{482,666} & \textbf{288} & \textbf{155,300} & \textbf{69,411} & 100 \\
\midrule

\multirow{2}{*}{C/C++} & \CW & 14.29 & 65.60 & 58,485 & 51.0 & 20,458 & 6,167 & 118 \\
    & \Sys & \textbf{27.06} & \textbf{71.95} & \textbf{501,823} & \textbf{255} & \textbf{117,761} & \textbf{70,946} & \textbf{159} \\
\midrule

\multirow{2}{*}{PHP} & \CW & 54.97 & 77.07 & 56,991 & 53 & 19,462 & 7,714 & 105 \\
    & \Sys & \textbf{68.21} & \textbf{78.48} & \textbf{275,304} & \textbf{142} & \textbf{80,133} & \textbf{41,635} & \textbf{125} \\
\midrule

\multirow{2}{*}{Average} & \CW & 40.09 & 67.16 & 74,032 & 56 & 26,374 & 9,190 & \textbf{157} \\
    & \Sys & \textbf{53.13} & \textbf{74.12} & \textbf{396,092} & \textbf{214} & \textbf{111,654} & \textbf{58,150} & 128 \\

\bottomrule
\end{tabular}
}
\end{table}

We analyze the results from three perspectives: \MetCov, \MetDI, and \MetComplK.

\textbf{\MetCov:} On average, \Sys achieves 53.13\% API coverage on multi-language evaluation, outperforming \CW (40.09\%) by 32.5\%. Notably, \RA achieves higher coverage (91.28\%) on Python, but this stems from its design goal of generating comprehensive API documentation that lists every referenced function. In contrast, \Sys aims to produce high-quality, high-level architecture and core module documentation with modular structure and cross-references. These represent different documentation philosophies: \RA prioritizes API completeness akin to traditional API docs, while \Sys prioritizes architectural clarity and semantic organization. As we show below, high coverage alone does not translate to better documentation quality or completeness for understanding system architecture.

\textbf{\MetDI:} Compared to \CW (Average), \Sys produces documentation with 5.3× more words (396K vs 74K), 3.8× more documented files (214 vs 56), 4.2× more cross-references (112K vs 26K), 6.3× more code blocks (58K vs 9K), and comparable diagrams (128 vs 157). On Python, \Sys significantly outperforms \RA in documentation structure, generating 1.4× more documented files (151.8 vs 110.7), 30× more code blocks (37K vs 1.2K), and actual diagrams (72.5 vs 0), despite \RA's higher word count from exhaustive function listing.

\textbf{\MetComplK:} On multi-language evaluation, \Sys achieves 74.12\% \MetComplK, outperforming \CW (67.16\%) by 10.4\%. On Python, \Sys (78.92\%) also outperforms \RA (76.53\%) by 2.4\%. \Sys shows particularly strong advantages in Java (+10.5\% vs \CW) and C/C++ (+6.4\% vs \CW). This demonstrates that despite \RA's high coverage, \Sys's documentation provides better information completeness to answer user queries.

\finding{blue}{\Sys achieves higher \MetCov (+32.5\% vs \CW) and \MetComplK (+10.4\% vs \CW, +2.4\% vs \RA), with significantly richer documentation content and structure.}

\subsection{RQ2: Documentation Readability and Richness}
\label{rq2}

This research question evaluates the documentation readability and richness of \Sys compared to baselines.

\begin{figure}[t]
\centering
\includegraphics[width=1\linewidth]{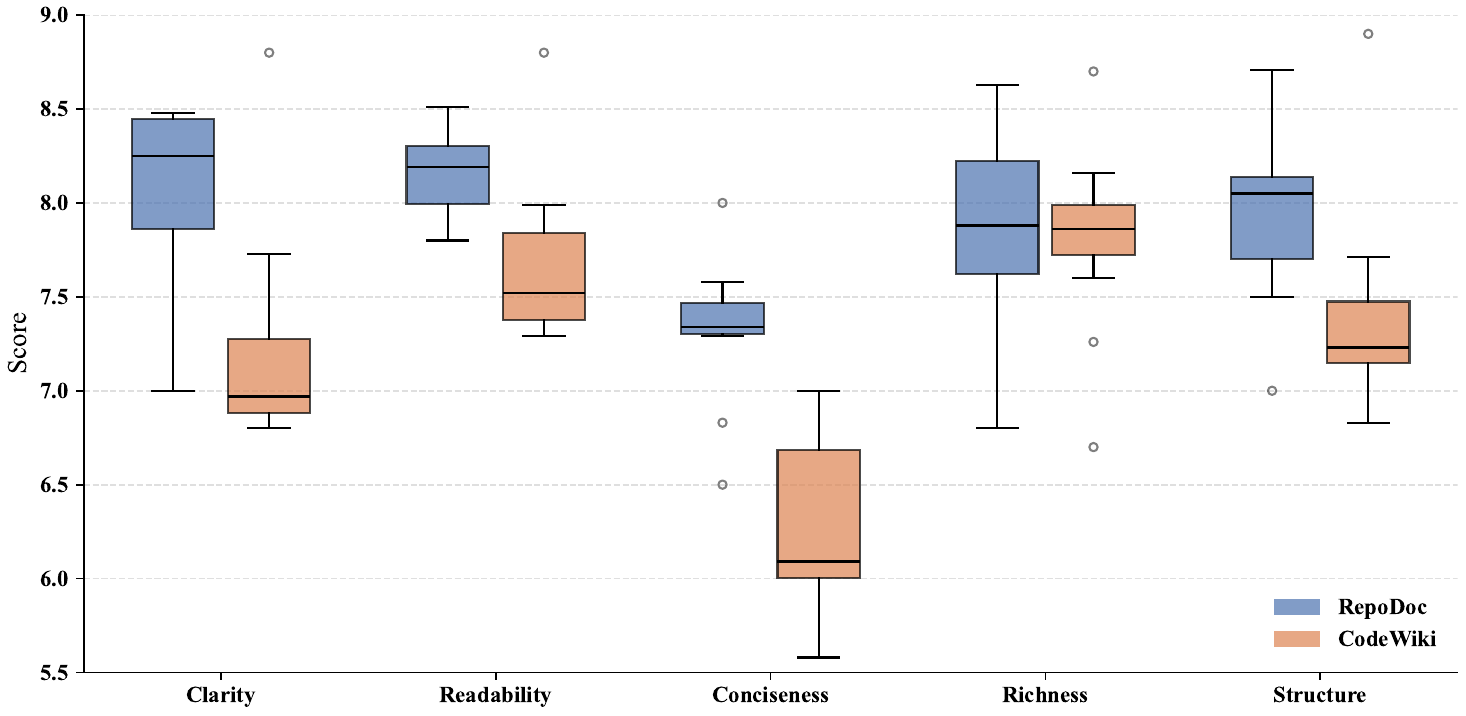}
\caption{\MetTQS box plot comparison}
\label{fig:tqs-boxplot}
\end{figure}

\subsubsection{Design}
To answer RQ2, we use \MetTQS (defined in Section \ref{benchmark}) to evaluate documentation readability and richness through pairwise comparison. For each repository in our dataset, we present LLM judges with documentation generated by both \Sys and \CW.

We exclude \RA from this evaluation for two reasons: (1) RQ1 has already demonstrated that \CW significantly outperforms \RA in documentation quality (e.g., 81 diagrams vs 0 diagrams), making the comparison redundant; and (2) RQ2 focuses on comparing the two state-of-the-art multi-language documentation tools, where the quality comparison between \Sys and \CW is more informative for understanding \Sys's positioning in the literature.

We aggregate scores across repositories where both tools generate documentation (i.e., Python, JS/TS, Java, C\#, C/C++, and PHP as shown in Table \ref{tab:method-comparison}) and visualize the distribution using box plots, as shown in Figure \ref{fig:tqs-boxplot}.

\begin{figure}[t]
  \centering
  \includegraphics[width=1\linewidth]{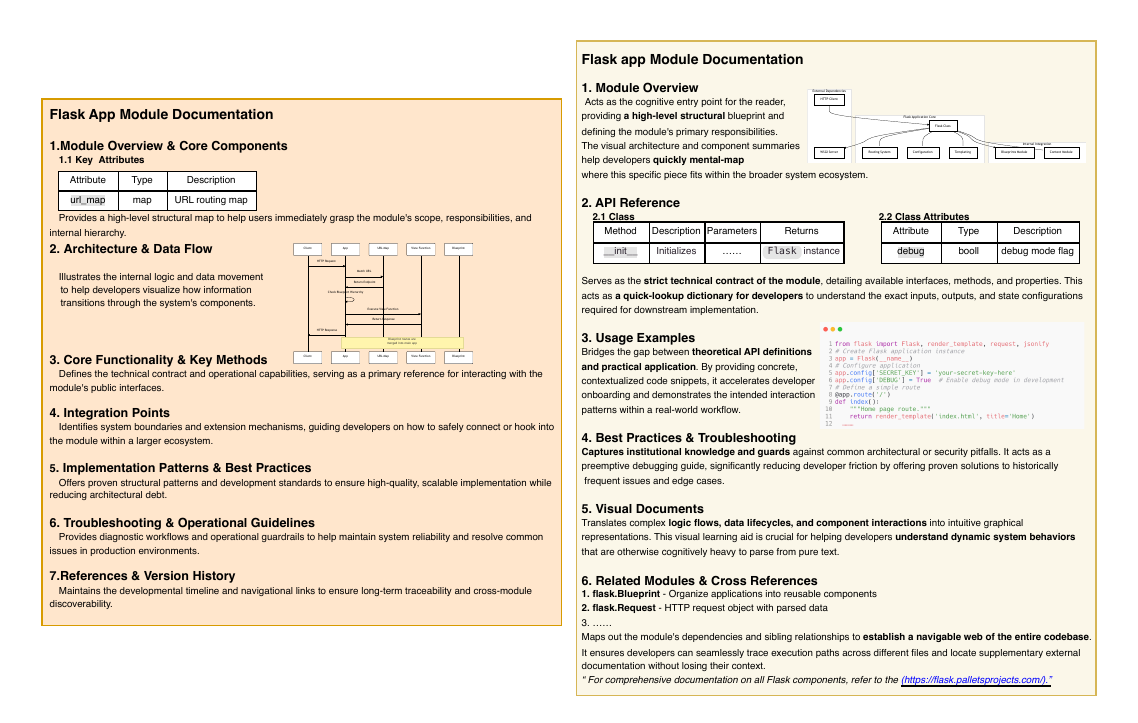}
  \caption{Qualitative comparison: \CW~(left) vs \Sys~(right)}
  \label{fig:doc-comparison}
\end{figure}

\subsubsection{Results}
Figure \ref{fig:tqs-boxplot} presents the \MetTQS comparison across five dimensions. \Sys outperforms \CW on all five dimensions. The advantage is most pronounced in Clarity (8.25 vs 7.00, +1.25) and Conciseness (7.40 vs 6.10, +1.30), followed by Structure (8.05 vs 7.20, +0.85) and Readability (8.20 vs 7.55, +0.65). The difference is minimal in Richness (7.90 vs 7.85, +0.05).

The observed advantages in Clarity, Conciseness, and Structure can be attributed to \Sys's \KG-based design. By extracting only relevant code snippets for context, \Sys avoids the context bloat that affects \CW's documentation, resulting in more focused and clearer explanations. The Mermaid diagrams generated by \Sys contribute to its superior Structure scores, providing readers with visual representations of system architecture. \CW's higher variance in Conciseness suggests that without a structured retrieval mechanism, its documentation tends to be inconsistent in balancing detail with brevity.

We manually reviewed the LLM-generated reasoning for each pairwise comparison. The primary reasons for \Sys's superiority are: (1) \KG-based context retrieval produces focused explanations with less redundancy; (2) Hierarchical module structure with clear cross-references improves navigation; (3) Auto-generated Mermaid diagrams provide visual architecture overviews. \CW's main weaknesses are that its explanations are relatively complex with more redundancy and fewer references and connections.

Figure \ref{fig:doc-comparison} illustrates a qualitative comparison between \CW and \Sys on the same code unit. \CW's documentation exhibits verbose explanations without clear modular organization, while \Sys provides structured modules with cross-references and architecture diagrams, demonstrating \Sys's superior Clarity and Structure as reflected in the \MetTQS evaluation.

\finding{green}{\Sys significantly outperforms \CW in Clarity (+1.25) and Conciseness (+1.30), with moderate advantages in Structure (+0.85) and Readability (+0.65), demonstrating superior overall documentation quality.}
\subsection{RQ3: Time and Token Efficiency}
\label{rq3}

This research question evaluates the efficiency of \Sys in terms of time and token consumption compared to baselines.

\begin{table}[t]
\small
\caption{Running Time and Cost Comparison (Time in seconds, Tokens in thousands, Cost in USD; DeepSeek V3.2, Python repos).}
\label{tab:rq1-cost}
\centering
\resizebox{1\linewidth}{!}{
\begin{tabular}{cccc}
\toprule
\textbf{Method} & \textbf{Time (s)} & \textbf{Tokens (K)} & \textbf{Cost (\$)} \\
\midrule

\Sys & 1,242.00 & 780.42 & 0.39 \\
\CW & 3,672.62 (+196\%) & 5,311.04 (+580\%) & 2.66 (+582\%) \\
\RA & 18,628.87 (+1400\%) & 4,704.05 (+503\%) & 2.35 (+503\%) \\

\bottomrule
\end{tabular}
}
\end{table}

\subsubsection{Design}
To answer RQ3, we measure efficiency in terms of running time and token consumption for all tools on Python repositories (the common ground for all three tools). We use DeepSeek V3.2 as the generation backbone for fair comparison. Token costs are converted to USD using DeepSeek V3 API pricing (\$0.5 per million tokens). The evaluation metrics include: \textbf{Time} measuring end-to-end documentation generation duration; \textbf{Token} measuring total tokens consumed during generation; and \textbf{Cost} calculated as Token consumption multiplied by the API pricing.

\subsubsection{Results}
Table \ref{tab:rq1-cost} shows the running time and cost comparison (averaged over Python repositories). \Sys achieves fast generation speed and low cost by leveraging the \RepoKG for context retrieval.

\Sys generates documentation in just \textbf{1,242 seconds} on average, compared to \CW (3,673s) and \RA (18,629s). In terms of cost, \Sys uses only \textbf{\$0.39} per repository, compared to \CW (\$2.66) and \RA (\$2.35). Specifically, \CW exceeds \Sys by 196\% in time, 580\% in tokens, and 582\% in cost, while \RA exceeds \Sys by 1,400\% in time, 503\% in tokens, and 503\% in cost. These savings stem from \Sys's \KG-based retrieval that extracts only relevant code snippets, avoiding the context bloat observed in baselines where entire files are fed into prompts.

\finding{purple}{\Sys significantly reduces generation time and cost compared to both baselines: \CW exceeds \Sys by 196\% (time), 580\% (tokens), and 582\% (cost), while \RA exceeds \Sys by 1,400\% (time), 503\% (tokens), and 503\% (cost). These savings stem from \KG-based retrieval that extracts only relevant code snippets.}

\subsection{RQ4: Incremental Update Performance}
\label{rq4}

This research question evaluates \Sys's capability in incremental updates compared to baselines.

\begin{table}[t]
\small
\caption{Incremental Update Comparison (DeepSeek V3.2, 20 scenarios per repository). Time in seconds, Token in thousands, Upd Recall in percent. $\checkmark$: used; $\times$: not used.}
\label{tab:incremental}
\centering
\resizebox{1.0\linewidth}{!}{
\begin{tabular}{cccccccc}
\toprule
\multirow{2}{*}{\textbf{Method}} & \multicolumn{2}{c}{\textbf{Info Source}} & \multirow{2}{*}{\textbf{Time (s)}} & \multirow{2}{*}{\textbf{Token (K)}} & \multirow{2}{*}{\textbf{Upd Recall}} \\
\cmidrule(lr){2-3}
    & \textbf{Base} & \textbf{Commit} &  &  &  &  \\
\midrule

\multirow{3}{*}{\RA} & $\checkmark$ & $\times$ & 18,629 & 4,704 & - \\
& $\checkmark$ & $\checkmark$ & 19,372 & 4,932 & 80.0 \\
& $\times$ & $\checkmark$ & 6,520 ($\downarrow$65\%) & 1,645 ($\downarrow$65\%) & 86.0 ($\uparrow$7.5\%) \\
\midrule

\multirow{3}{*}{\Sys} & $\checkmark$ & $\times$ & 1,242 & 780 & - \\
& $\checkmark$ & $\checkmark$ & 1,315 & 835 & 88.0 \\
& $\times$ & $\checkmark$ & \textbf{342} (\textbf{$\downarrow$73\%}) & \textbf{181} (\textbf{$\downarrow$77\%}) & \textbf{97.0} (\textbf{$\uparrow$10.2\%}) \\

\bottomrule
\end{tabular}
}
\end{table}

\subsubsection{Design}
We evaluate \Sys's incremental update mechanism on Python repositories from \Bench. We design three documentation generation scenarios: \textbf{Base} represents the initial full documentation generated from the code state \textit{before} the selected commit; \textbf{Full} denotes full regeneration from scratch at the new commit; and \textbf{Update} refers to incremental updates that modify only affected documentation components rather than regenerating everything. For each repository, we first identify 20 real commits covering diverse change patterns (API additions/modifications, feature implementations, bug fixes, refactoring, 5 each). For each selected commit, we generate Base documentation from the code state immediately before that commit, then apply the commit changes and evaluate both Full regeneration and incremental Update. We compare against \RA as the baseline since \CW does not support incremental updates. The evaluation metrics include Time, Token consumption, and Upd Recall (defined in Section \ref{benchmark}). To properly assess incremental update capability, we measure the \textit{recall improvement rate}---the difference between Update and Base recall---rather than absolute values, as this captures how effectively a system identifies affected components during code evolution.

\subsubsection{Results}
Table \ref{tab:incremental} presents the incremental update comparison. \Sys's incremental update mechanism leverages \SIP through the \RepoKG, reducing update time by \textbf{73\%} and token usage by \textbf{77\%} compared to full regeneration. More importantly, incremental updates with \Sys achieve \textbf{10.2\% higher Upd Recall} (defined in Section \ref{benchmark}) than full regeneration, showing that the system can identify and incorporate code changes more effectively. This improvement is attributed to \Sys's \KG-based \SIP, which precisely identifies all affected components through dependency traversal.

\textbf{Understanding the Upd Recall metric.} The \textit{Upd Recall} metric measures how well the updated documentation captures the code changes, computed as the ratio of correctly updated components to total components requiring update. \RA achieves 86.0\% Upd Recall with 7.5\% improvement over full regeneration (80.0\%). In contrast, \Sys achieves 97.0\% Upd Recall with 10.2\% improvement over full regeneration (88.0\%), demonstrating that \SIP through the \RepoKG can accurately track code changes across module boundaries.




\finding{orange}{\Sys's \KG-based \SIP enables effective incremental documentation updates: it reduces update time by 73\% and token usage by 77\% through selective regeneration. \Sys achieves a 10.2\% Upd Recall improvement through incremental updates, compared to \RA's 7.5\% improvement, demonstrating that \KG-based semantic traversal precisely identifies affected components when code changes.}

\section{Related Work}
\label{related}

Our work relates to three research areas: code documentation generation, knowledge graphs in software engineering, and benchmarks for code documentation generation.

\textbf{Code Documentation Generation}. Early work relied on rule-based or template-based approaches for generating comments~\cite{10.1145/1858996.1859006}, which were often limited in expressiveness and struggled to generalize across diverse codebases. Deep learning models later enabled more natural comment generation~\cite{ahmad2020transformerbasedapproachsourcecode,iyer-etal-2016-summarizing}, improving fluency and semantic alignment between code and generated descriptions. The LLM era has advanced repository-level documentation: \CW~\cite{makharev2025codewiki} leverages retrieval-augmented generation, while \RA~\cite{luo2024repoagent} employs iterative refinement with bidirectional reference graphs, alongside other multi-agent or retrieval-based frameworks~\cite{zhang-etal-2023-repocoder,yang2024sweagent}. However, both treat code as flat text chunks without semantic modularity, ignoring higher-level structural and dependency information inherent in software systems. \textbf{\Sys constructs a \RepoKG capturing dependencies, enabling structurally coherent documentation with cross-references and Mermaid diagrams.}

\textbf{Knowledge Graphs in Software Engineering}. Knowledge graphs have been applied to API knowledge bases, call graphs~\cite{10.1145/1595696.1595767,chen2022apiusagerecommendationmultiview}, and pre-trained models like GraphCodeBERT for code search and bug detection~\cite{guo2021graphcodebertpretrainingcoderepresentations}, demonstrating their effectiveness in capturing structural and semantic relationships in code. \textbf{No prior work applies knowledge graphs to the full documentation lifecycle, particularly for both generation and maintenance. \Sys fills this gap by leveraging \RepoKG as the semantic backbone to precisely identify, organize, and update affected documentation components throughout the lifecycle.}

\textbf{Benchmarks for Code Documentation Generation}. Existing benchmarks like CodeSearchNet~\cite{husain2020codesearchnetchallengeevaluatingstate} and CodeXGLUE~\cite{lu2021codexgluemachinelearningbenchmark} focus on function-level metrics. Recent work like CodeWikiBench evaluates repository-level documentation but lacks assessment of incremental updates. Other recent evaluations, like Long Code Arena ~\cite{bogomolov2024longcodearenaset}, address large-scale repositories but focus on input length and scalability challenges rather than documentation lifecycle management. \textbf{\Sys introduces \Bench with five evaluation dimensions including \MetUR, providing the first systematic framework to evaluate incremental documentation updates alongside generation quality and efficiency.}

\section{Threats to Validity}
\label{threats}

We discuss potential threats to validity concerning our experimental evaluation.

\textbf{Construct Validity}. We ensure our five evaluation dimensions accurately measure documentation quality. \MetCov measures API documentation presence; \MetComplK assesses whether documentation answers user queries; \MetTQS uses LLM judges for pairwise quality comparison across Clarity, Readability, Conciseness, Richness, and Structure; \MetUR measures incremental update effectiveness; and \MetDI evaluates richness across multiple structural elements. We use multi-model LLM judges and fixed prompts to reduce bias, and release our evaluation protocols for reproducibility.

\textbf{Internal Validity}. We address bias from favorable repository selection by evaluating on 24 repositories across 8 languages and 3 scales; RQ2 uses pairwise comparison on repositories where both tools generate docs; RQ3 measures efficiency on Python where all tools run; and RQ4 evaluates 20 real commits per repository. For fair baseline comparison, we use default settings and identical hardware.

\textbf{External Validity}. While we evaluate across diverse repositories, performance may vary on private codebases or domain-specific applications. \Bench enables future researchers to assess generalizability under identical metrics. We release implementation code and datasets to facilitate independent verification.

\section{Conclusion}
\label{conclusion}

This paper presented \Sys, an automated documentation generation system that leverages \RepoKG as the backbone for the entire documentation lifecycle. The key insight is that a \RepoKG built once can be traversed to precisely identify affected components when code changes, eliminating the need for full regeneration. Our three-phase pipeline produces structurally coherent documentation with automatic cross-references and Mermaid diagrams, while \SIP enables efficient incremental updates. Experiments on 24 repositories across 8 programming languages demonstrate that \Sys significantly outperforms state-of-the-art baselines in both documentation quality and update efficiency. We hope this work inspires future research on \KG-based software engineering tools for automated documentation maintenance.



\bibliographystyle{ACM-Reference-Format}
\bibliography{sample-base,software}

\end{document}